# Title: The Paracrystalline Nature of Lattice Distortion in a High Entropy Alloy

**Authors:** Yu-Tsun Shao[1], Renliang Yuan[1], Yang Hu[1], Qun Yang[1], and Jian-Min Zuo[1,2*]

**Affiliations:**

[1]Department of Materials Science and Engineering, University of Illinois at Urbana-Champaign, Urbana, Illinois 61801, USA.

[2]Frederick Seitz Materials Research Laboratory, University of Illinois at Urbana-Champaign, Urbana, Illinois 61801, USA.

*Correspondence to: jianzuo@illinois.edu.

**One Sentence Summary:** Electron nanodiffraction reveals the paracrystalline structure of a high entropy alloy and its complex strain field.

**Abstract:** Severe lattice distortion is suggested for high entropy alloys (HEAs), however, evidence for such effect so far is lacking, and the nature of distortion is yet to be understood. Here, we reveal the distortion in an fcc HEA, $Al_{0.1}CrFeCoNi$, by direct imaging using electron nanodiffraction. Information about crystal symmetry, lattice strain and atomic distortion are data-mined and mapped from many ($\sim 10^4$) diffraction patterns. Application to the HEA reveals two embodiments of distortion, nm-sized mosaic blocks of paracrystals and strained nano-clusters. Their interaction gives rise to fractal strain field across nanoscopic to mesoscopic scales. As lattice distortion impedes dislocation motion and contributes to strengthening, results here thus provide critical insights about the complex nature of distortion in a HEA.

## Introduction:

High entropy alloys have attracted tremendous research attention recently (*1-5*). These alloys are chemically concentrated, comprised of at least five elements in an equal or near-equal atomic fraction, and yet they can be structurally "simple" as single-phase, disordered, solid solution. The stabilization of the solution phase can be attributed to the configurational entropy of mixing, which is maximized in the HEAs (*6*). In a solid solution, atoms are randomly distributed among the available sites in a crystal lattice. The mismatch in the atomic radius of different elements would then result in a large number of atoms being displaced from their ideal positions (lattice distortion) and stressed (atomic stress). Consequently, severe lattice distortion has been suggested as one of core effects in HEAs (others are aforementioned entropy of mixing, sluggish diffusion and cocktail effect) (*6, 7*). In a conventional alloy, lattice distortion and the related stress field impede dislocation motion, leading to solid-solution strengthening (*8-10*). Similarly, lattice distortion has been suggested for the pronounced strengthening observed in the HEAs (*11-13*).

However, unlike conventional alloys, the distinction between the solute and matrix atoms disappear in the HEAs. The complex chemical and lattice disorder in the HEAs present a major challenge for their characterization (*7*). Traditionally, crystal disorder is studied by diffraction based on the analysis of Bragg peak broadening (*14, 15*), Debye-Waller factor (*16*), diffuse scattering patterns (*17*) and pair-distribution function (PDF) (*18*). The analysis is performed over a volume of crystals. So far, such studies provided inconclusive evidence for lattice distortion (*19, 20*). For example, PDF analysis of atomic distances in CrMnFeCoNi and Ni alloys demonstrated no significant differences between the HEA and the conventional alloys (*21*).

For characterizing structural disorder in HEAs, two basic kinds of lattice distortion must be distinguished, with the presence (1st) and absence (2nd) of the long-range crystal order (*14, 15*). In the 1st kind, a solid-solution alloy has an average lattice where atoms are displaced from the lattice sites. While in the 2nd kind, the lattice is distorted, the distortion fluctuates from one site to another, and the fluctuation increases as the distance to the site increases. Such structure is described as paracrystal (*14, 22-24*). The models of paracrystal have been applied to describe the intermediate structures formed between perfect crystals and amorphous structure in a broad range of materials, including alloys, colloidal suspensions, organic semiconductors, polymer and protein crystals (*15*). Support for the paracrystal model was also provided by electron microscopy (*25, 26*). However, the observed paracrystals often have a mixture of ordered and disordered phases, and the interplay between order and disorder in a paracrystalline alloy is not clear. Real alloys also have defects and the related strain field, which must be separated as well in the study of lattice distortion (*20*).

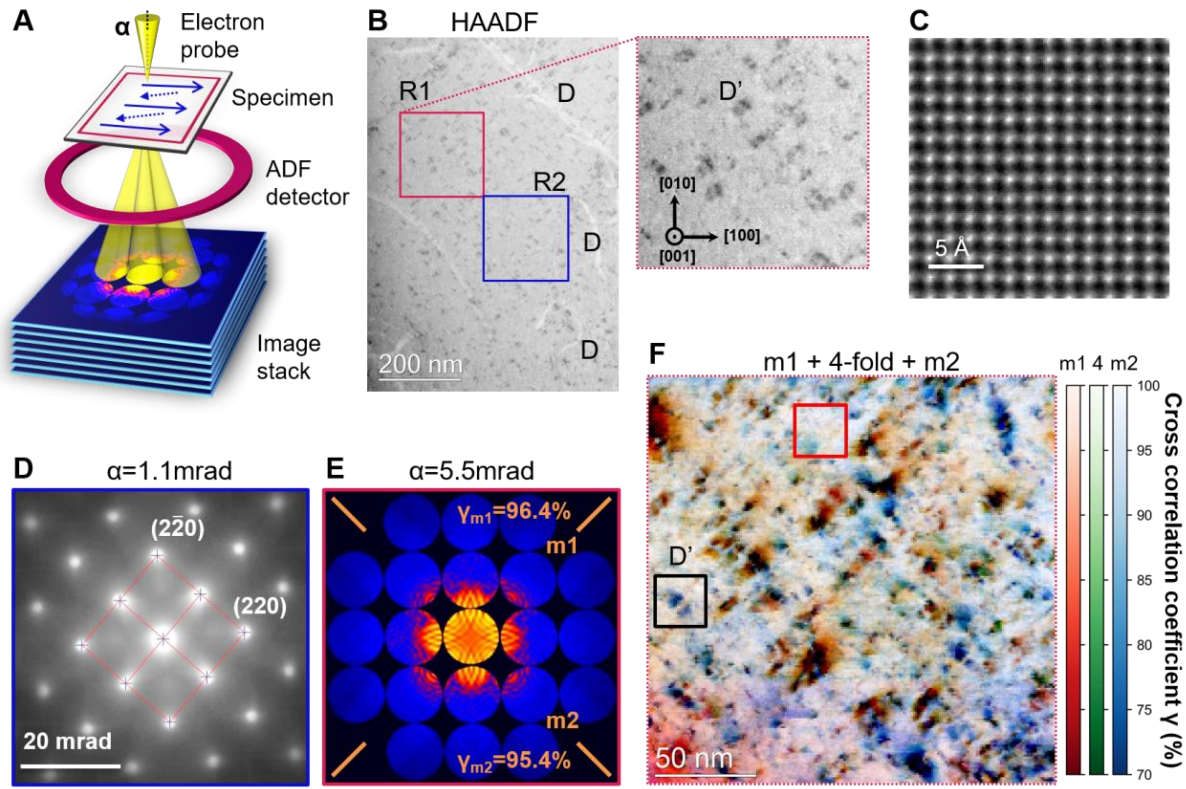

***Fig. 1. Imaging of distorted lattice in the HEA of Al$_{0.1}$CrFeCoNi.*** *(**A**) Schematic of the STEM and SEND imaging using a scanning probe of semi-convergence angle α. SEND records a diffraction pattern at each probe position, while STEM signal is collected by the annular dark field (ADF) detector. (**B**) HAADF image acquired with α = 5.5mrad, inner collection angle of 60 mrad. R1 and R2 indicate the regions of the SEND experiments were performed. D and D' marks the*

*observed defects. (**C**) Atomic resolution STEM image acquired with α = 25 mrad. (**D**) Electron nanodiffraction pattern formed with α = 1.1mrad, strain can be obtained by measuring relative shifts in the disk positions. (**E**) Representative experimental CBED pattern with α = 5.5 mrad, having highest symmetry of $\gamma_4$=95.9%, $\gamma_{m1}$=96.4%, $\gamma_{m2}$=95.4%. (**F**) Overlay of symmetry maps obtained by quantifying symmetry in CBED patterns for 4-fold rotation (green), and along two different mirror plane directions, m1 (red) and m2 (blue). With the aid of CBED simulations, the specimen thickness is determined as ~186nm.*

Here, we report on a scanning electron diffraction (SEND) based approach for the determination of distortion fields, and its application to an fcc HEA, $Al_{0.1}CrFeCoNi$ alloy. SEND works by rastering a nm-sized focused electron probe over a selected sample area, and by recording electron diffraction patterns over each probe position using a pixelated detector (Fig. 1A, also see Ref. (*27*)). Inside a transmission electron microscope (TEM), SEND analysis can be carried away from dislocations, grain boundaries or other bulk defects. While the principles of SEND is similar to the recently reported 4D-STEM technique (*28*), a key difference is that the small beam convergence angle in SEND enables quantitative measurement of crystal symmetry, lattice strain, and distortion. Using such approach, we show that the structure of the HEA is comprised of nm-sized mosaic blocks of paracrystals. Strain is found between mosaic blocks and in nm-sized strained clusters, and their interaction gives rise to the inhomogeneous strain fields with fractal patterns. The inhomogeneous strain field is related to the nanoscale dislocation dynamics observed in the HEA (*29*).

**Experimental methods**

The SEND experiments here were primarily carried out using a Themis Z STEM (scanning TEM, Thermo-Fisher Scientific, operated at 300kV). The polycrystalline $Al_{0.1}CrFeCoNi$ sample was prepared and characterized as described in Ref. (*29*). Thin TEM specimens were prepared along [001] using the focused ion beam (FIB) method (*30*). (Further details are provided in Supplementary Information.) Figures 1B and C show two STEM images obtained at 0.8 nm and sub-Å spatial resolution. Regular fcc lattice is observed in Fig. 1C, while Figure 1B reveals two types of contrast, the bright lines are dislocations in the as-prepared sample, and the dark coffee-bean-like contrast. Diffraction analyses were performed in two regions of the HEA sample away from dislocations (marked in Fig. 1B), using two types of diffraction patterns of nanodiffraction (Fig. 1D) and convergent beam electron diffraction (CBED, Fig. 1E). The CBED patterns were used to determine local crystal symmetry. Electron nanodiffraction was used for the measurement of local *d*-spacing, lattice strain, and paracrystalline distortion.

**Results**

Figure 1F shows the HEA structure as measured by the CBED pattern symmetry. A perfect fcc crystal is expected to exhibit the *4mm* symmetry in CBED along the cubic axis. The measured symmetries in Fig. 1F include two mirrors (*m1* and *m2*, as marked in Fig. 1E), plus the 4-fold rotation. The pattern symmetry was quantified by calculating the correlation coefficient ($\gamma$) between the symmetry related diffraction disks, following the procedures described in refs. (*31, 32*). For a pattern of perfect symmetry, $\gamma$ = 100%, while the measured $\gamma$ in the single crystals of Si is close to 98% (*31, 32*). In comparison, the highest measured $\gamma$ in the HEA is 93%. In Fig. 1F, in regions where the HAADF-STEM image contrast is relative uniform, the symmetry of CBED patterns fluctuates between 83-93%. Large symmetry breaking is observed around defects observed in STEM (marked as D' in Fig. 1B). Around these defects, $\gamma$ ranges from 43 to 70%.

Figure 2A highlights a region where the γ fluctuates. The symmetry fluctuations can be attributed to small random displacements seen by the electron beam. To demonstrate this and also to estimate the magnitude of these random displacements, we performed a series of CBED simulations using a mosaic block model (Fig.2B). In this model, the crystal column is divided into nm-sized blocks, each block is displaced by ($R_x$, $R_y$, $R_z$), which are random and follow a Gaussian distribution of width $\sigma_R$ and zero mean (Fig. 2B). We simulated a total of 18,125 CBED patterns with different mosaic block configurations and quantified their symmetry. The simulations covered a range of block size ($L_b$) and the standard deviation $\sigma_R$ of displacements, with $L_b$= 1-10 nm and $\sigma_R$ = 0.013-0.023 Å, respectively. The simulation results were compared with the distribution of 4-fold rotational symmetry of 4,375 experimental CBED patterns selected from regions free of inclusions. Our results show that the local symmetry fluctuation can be approximated by 3-8 nm sized blocks, which are displaced with the standard deviation of $\sigma_R$ = 0.02 Å (for details, see Fig. S1). For example, Fig. 2D shows a simulated CBED pattern of $L_b$ = 5 nm and $\sigma_R$ = 0.02 Å, which corresponds approximately to the experimental CBED pattern shown in Fig. 2C.

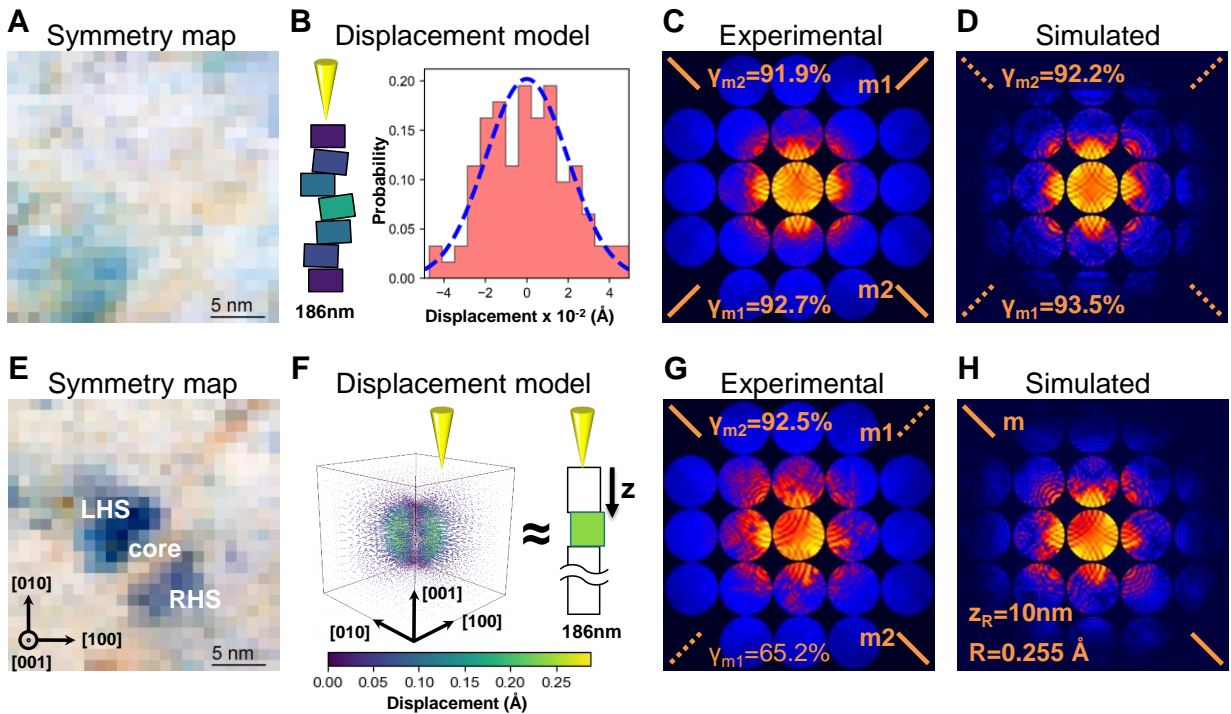

*Figure 2. Two types of lattice distortions in Al$_{0.1}$CrFeCoNi HEA observed by CBED. The two are distinguished by the symmetry maps of (A) and (E) from the colored boxes in Figure 1F, and they are interpreted by the models of (B) and (F), respectively. The displacements due to a distorted lattice along the electron beam path are approximated as a column having (F) a block with a large displacement as color coded, and (B) small blocks with displacements following the Gaussian distribution. The displacement model in (F) is an approximation of the fields from Eshelby's theory of ellipsoidal inclusions (33). (D and H) show the simulated CBED patterns corresponding to the experimental patterns in (C and G), respectively. CBED patterns from imperfect crystals were simulated using the Bloch-wave scattering-matrix method as described in Supplementary Text.*

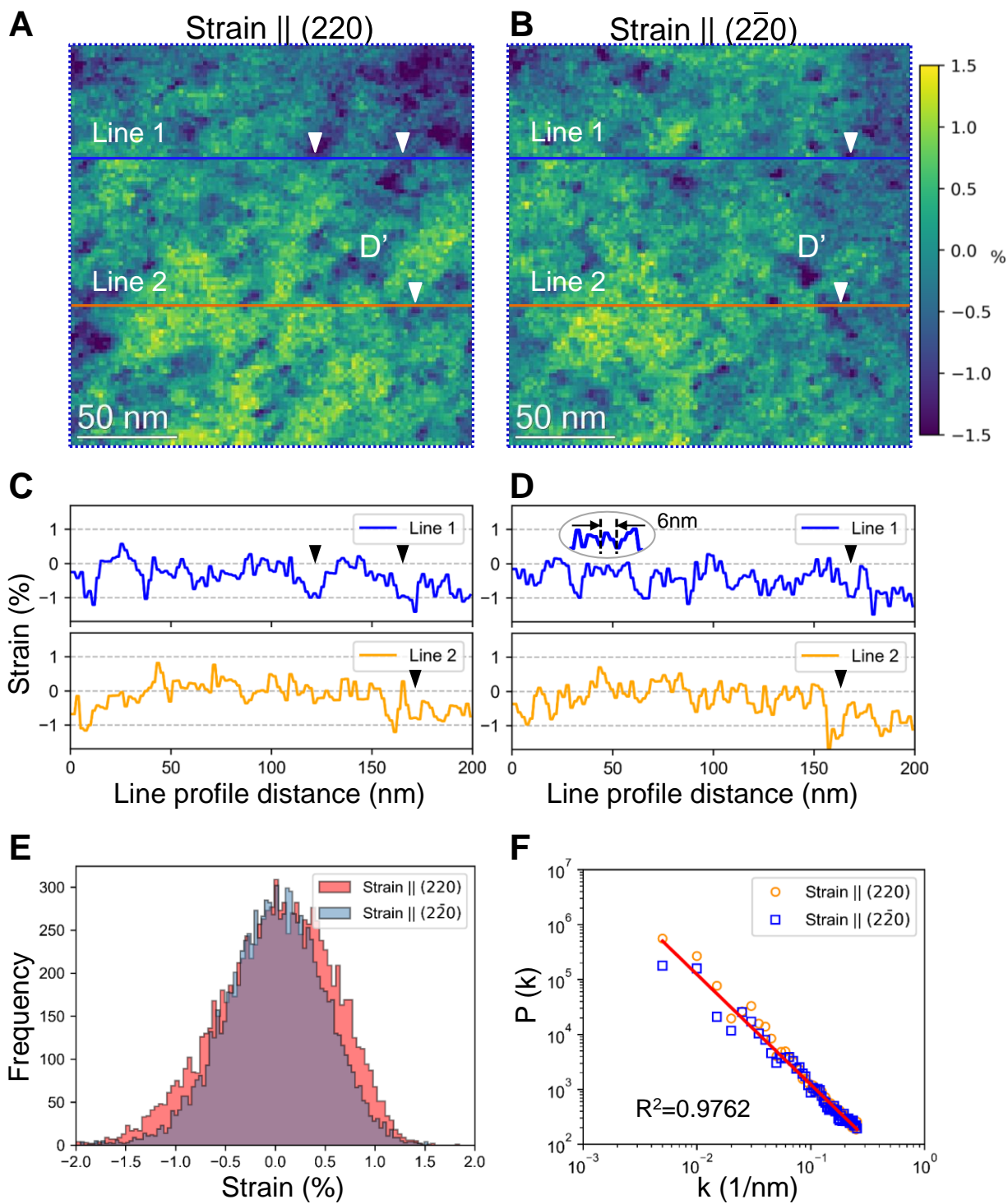

**Figure 3. Lattice strain and strain fluctuations in Al₀.₁CrFeCoNi HEA.** *Strain maps of region R2 along the (**A**) (220) and (**B**) (2̄20) directions were obtained by measuring relative shifts in disk positions. The maps were obtained from region R2 of Fig. 1B using a 100x100 scan over 200x200 nm² with an electron probe of 1.7 nm in full-width at half-maximum (FWHM). (**C** and **D**) Line profiles drawn from (A and B), respectively. The triangles mark regions with significant*

*compressive strain, which correspond to inclusions. (**E**) Histograms showing the distribution of lattice strain along (220) and (2$\bar{2}$0). (**F**) The scaling of P(k) versus spatial frequency k, where P is the square of Fourier amplitude of strain maps in (A) and (B).*

The defect regions of D' with large symmetry breaking are examined in Fig. 2. These regions feature anisotropic symmetry breaking along <110> directions, and a core, the left-hand-side (LHS) and the right-hand-side (RHS) regions, as the example shown in Fig. 2E. The CBED patterns from these regions show different symmetry. For example, the experimental CBED pattern (Fig. 2G) taken from RHS of Fig. 2E shows a higher mirror symmetry along *m2* than *m1*, with $\gamma_{m2}$=92.5% and $\gamma_{m1}$=65.2%, respectively. The symmetry breaking is in close resemblance of the projected normal stress field around an inclusion of an oblate spheroid, as calculated from the Eshelby's model for ellipsoid inclusions (Fig. 2F). The ellipsoid resides in the (110) plane in Fig. 2F. To see the impact of displacements from the ellipsoid, we simulated CBED in Fig. 2H by varying the displacement ($\vec{R}$) and other parameters (see Supplementary Text on Simulations). A good match with the experimental pattern (Fig. 2G) was obtained with $|\vec{R}|$=0.254 Å ($|\vec{R}|/a_0$=7.1%, $a_0$=3.583Å is the average lattice constant) and $\vec{R}$||(2$\bar{2}$0).

The inclusions are relatively common as Fig. 1B shows. Within Region R1 in Fig. 1B, 22 such inclusions were identified, and their average dimension is found to be 10±3 nm in diameter. All 22 inclusions show displacements away from the core along one of the <110> directions.

Next, we mapped the lattice strain by measuring the d-spacing of (220) and (2$\bar{2}$0) reflections and used them to calculate the lattice strain along these directions (Fig. 3). The measurement of d-spacing follows the methods described in Supplementary Text. A measurement precision (±0.13%) was achieved (Fig. S2). For the HEA, the measured strain values in Figs. 3A and 3B have the standard deviation of 0.58% and 0.49%, respectively, with the reference d-spacing taken as 1.267 Å. Different types strain variations are observed in Fig. 3. The symmetry-breaking inclusions observed in Fig. 1F correspond to regions with compressive strain in Figs. 3A and 3B (marked by filled white triangles). The largest compressive strain measured by SEND is -2.3% (compared to -7.1% estimated from CBED, the smaller value is expected since nanodiffraction measures the average strain over the sample thickness). Away from the inclusions, the strain maps reveal clusters of strained regions and the line profiles over these regions (Figs. 3C and D) show varying strain over different length scales. Figure 3F shows the correlation of strain by plotting the power spectrum P(k) as function of k (the spatial frequency *1/L*) for the sample region R2, where P(k) is radially averaged intensity of Fourier transform of the two strain maps in Fig. 3A and 3B. Results here demonstrates $P(k) \propto k^{-2}$ for k = 0.004 to 0.2 nm$^{-1}$.

To determine the nature of lattice distortion, we examined diffraction peak broadening in the recorded SEND patterns. The basic idea is that the 2$^{nd}$ kind of lattice distortion causes a nonlinear increase in the integral breadths $\delta b$ of successive order (*h*) of reflections, while the 1$^{st}$ kind lattice distortion causes a decrease in the intensities of the Bragg reflections only (*14, 15*). In real crystals, inhomogeneous strain also contributes to peak broadening, which is proportional to tan$\theta$ ($\theta$ for diffraction angle). The $\delta b$ is measured by the area under the diffraction peak divided by its height. The instrument effect was calibrated using a Si single crystal (Fig. S3), which gave a relatively constant $\delta b$ (~0.7$\delta b$ of the {200} reflections of the HEA). In Fig. 4B, the measured $\delta b$ of the {*h*00} (*h* = 2, 4, and 6) reflections are plotted as function of $h^2$, which is fitted by the nonlinear model (*34*)

$$\delta b = \frac{1}{\langle L \rangle} + \frac{(\pi g h)^2}{\langle d_{hkl} \rangle}, \quad (1)$$

where $\langle L \rangle$ and $\langle d \rangle$ are the average crystal size and d-spacing for the ($hkl$) plane, and $g$ is the paracrystalline distortion parameter (PDP). From Fig. 4B we obtained $\langle L \rangle = 1.5\ nm$, which is comparable to the electron probe size of 1.7 nm in FWHM. Figure 4C shows the map of $g$ obtained from the peak width analysis from region R2 (Fig. 1D). The mean lattice distortion parameter g in this region was determined as 2.47% with a standard deviation of 0.38%. An example of the fluctuations in $g$ is

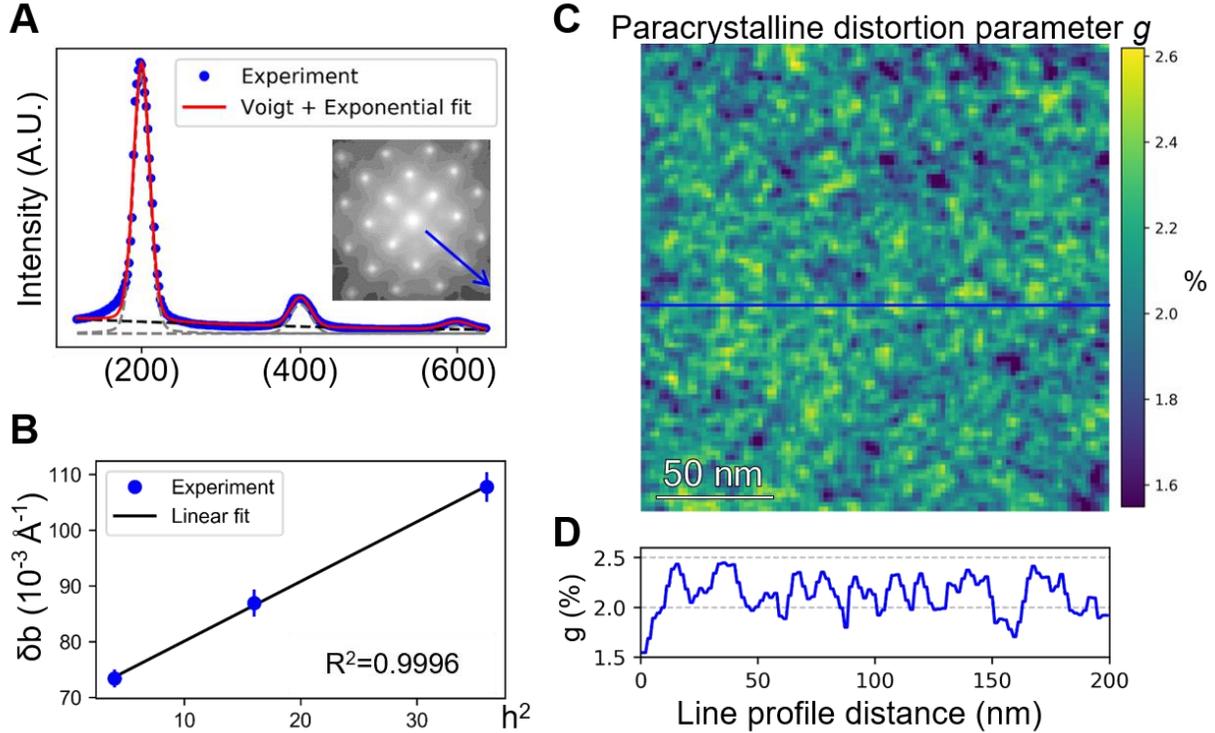

*Figure 4. Measurement and mapping of paracrystalline distortion parameter g in $Al_{0.1}CrFeCoNi$. (**A**) A fit to the intensity profile over three orders of the (h00) reflections taken from the averaged electron nanodiffraction pattern. (**B**) The integral breadth $\delta b$ is plotted versus $h^2$ and fitted by a line, whose slope gives g = 2.47%. The error bars represent the standard deviation of Si calibration (see Suppl. Info. for details). (**C**) The g map of region R2 of 100x100 pixels (1 pixel = 2 nm), obtained by measuring the peak broadening in the diffraction pattern of each pixel and after a Gaussian smoothing of 1 pixel width (2 nm). (**D**) A profile of (**C**) over the marked blue line.*

## Discussion

The above results demonstrated the paracrystalline nature of lattice distortion in the HEA of $Al_{0.1}CrFeCoNi$ and its inhomogeneous strain field. A paracrystal is characterized by the variance of *d*-spacing ( $g^2 = \langle d^2 \rangle / \langle d \rangle^2 - 1$ ). This variance differs from the mean-square atomic displacements $\langle u^2 \rangle$ (1st kind distortion) in the Debye-Waller factor. The correlated nature of lattice distortion as prescribed by $g^2$ can be understood using the perturbed lattice model of Eads and Milane (*24*) for a one-dimenional (1D) paracrystal. In this model, the position of *j*th lattice site is given by

$$\vec{r}_j = \vec{r}_{j-1} + \vec{d}_j = \sum_{k=1}^{j} \vec{d}_k \ , \ (2)$$

where the $\vec{d}_j$ are random vectors which have two components, along ($d_{j\parallel}$) and normal ($d_{j\perp}$) to the 1D crystal axis, and $\langle d_{j\parallel} \rangle = d_o$ (the average period of the 1D paracrystal) and $\langle d_{j\perp} \rangle = 0$. The $\vec{r}_j$ is dependent on the displacement of the sites of *1* to *j-1*. For a 1D paracrystal, the diffraction peak width $\delta b$ increases with increasing scattering angle, and follows approximately the relationship of eqn. 1 for low-order reflections in the perturbed lattice model (*24*). The diffraction pattern from a two-dimensional paracrystal is taken as equal to the product of the diffraction patterns from two 1D paracrystals (*15, 24*).

A consequence of the paracrystal model (*eqn.* 2) is an increase in the variance (var) as the lattice point moves further away from the origin, e.g., $\mathrm{var}\left(r_{j\parallel}\right) = \mathrm{var}\left(\sum_{k=0}^{j} d_{k\parallel}\right) = j \cdot \left\{\left\langle d^2 \right\rangle - \left\langle d \right\rangle^2\right\}$. Because of this, large fluctuations in the lengths of the unit-cell edges are predicted on the surface of the paracrystal, and thence to a much distorted cell, as the size of a paracrystal increases. Previous studies have established an empirical $\alpha^*$ law that limits the size of a paracrystal (*N*) to $\sqrt{N}g = \alpha^* = 0.15 \pm 0.05$ (*15, 35*). The $\alpha^*$ law implicitly indicates that in a paracrystal the crystallite size stops grow when the new cell to be added is too distorted. For $g = 2.4\%$, $N \approx 39$ and the predicted crystallite size is ~5 nm. In the HEA, our study shows that these crystallites manifest as paracrystalline domains or mosaic blocks. Fluctuations in the lattice strain are observed at the length scale similar to the size of the crystallites, which suggests an interruption to the paracrystalline order by strain relaxation. Such interruption is also evidenced by the symmetry fluctuations observed by CBED, which suggested a Gaussian distributed lattice displacements with $\sigma_R = 0.02$ Å between the mosaic blocks of 5 nm in size.

The above discussion demonstrates the role of the *g* in determining the structure of a HEA. The variance of *d*-spacing originates from the atomic size misfits and the local chemical composition. A parameter that has been proposed to estimate the size mismatch in a HEAs is the misfit factor $\delta = \sqrt{\sum_{i=1}^{N} c_i \left(1 - r_i / \sum_{j=1}^{N} c_j r_j\right)^2}$, where *N* is the number of elements in HEA, $c_i$ and $r_i$ are the atomic fraction and radius of *i*th element, respectively. The underlining model for $\delta$ assumes hard sphere packing. It has been shown that the $\delta$ factor is correlated to the atomic strain ($\varepsilon_A$) in the models of the HEA structure constructed based on density function theory (DFT) based calculations, with $\varepsilon_A \approx f(v)\delta$ and $f(v) \sim 1$ (*36*). The measured *g* with values ranging from 1.2-3.6% are comparable with the estimated $\delta$ of 1.2-4.8% in the $Al_xCrFeCoNi$ system for x = 0 - 0.15.

In regions with inclusions, important questions are raised about their origin and thermodynamic stability. The inclusions suggest the non-ideal solid-solution nature in $Al_{0.1}CrFeCoNi$. Previously, the spinodal decomposition of Cr, Fe-rich versus Al, Ni, Co-rich segregations was observed in the $Al_xCrFeCoNi$ systems at the higher *x* (*37*). First-principles electronic structure calculations demonstrated that the interactions between Al and transition metal elements are strongly attractive, Cr and Fe tend to stabilize the bcc structure and Ni and Co work as fcc stabilizer in the alloys, and the structures with ordered Al atoms at random transition metal atoms are more stable at higher Al concentrations (*38*). The inclusions we observed are disk-like in {110} planes and tensile-strained at the core, which can be assumed as Al-rich, since Al has the largest atomic-size mismatch among the constitutive atoms. Electron nanodiffraction from the core of inclusions shows strong diffuse scattering but no extra sharp reflections, which indicates short-range intermetallic ordering (Fig. S4). The Al-rich inclusions would be similar with the Pd-Pd pairs identified in PdCrFeCoNi, which also show the strain field following Eshelby's inclusion theory (*19*). The sizes of inclusions are rather uniform at ~10 nm in diameter. The small sizes suggest that strain energy counter-balances the trend toward Al segregation.

The presence of Al-rich inclusions and the fluctuations of the PDP demonstrate fluctuations in the local chemical composition at the length scales of few to tens of nm, respectively, which is another major characteristic of the HEA. This is consistent with the atom probe analysis (*29*) and EDS mapping (Fig. S5), as both measurements indicate fluctuations in chemical compositions at the nanoscale. The chemical fluctuations couple with the distortion fields of the paracrystals and the inclusions to create an inhomogeneous strain field. the distance-dependent strain correlations in the HEA follows a power law, with $P(k) \propto k^{-2}$ (Fig. 3) and the pattern observed in the strain map is fractal (Fig. S6). Fractal patterns are observed in numerous natural systems and in phase transformation. The presence of power law indicates the ability of a system to self-organize when constantly pushed to a disordered state by natural perturbations (*39*).

Together, the above results demonstrate a fractal distortion field in the HEA that contributes to the strengthening. Recently, Hu et al. reported dislocation storage and dislocation pileup that led to dislocation avalanches in the nanopillars of $Al_{0.1}CrFeCoNi$ using in-situ TEM, which they attributed to dislocation pinning by lattice distortion (*29*). Theory predicts higher alloy strength for large misfit, which is determined by the atomic misfit factor $\delta$ as well as its fluctuations (*10*). The findings here demonstrate the nature of the lattice distortions and thus insights into the strengthening in the HEA. Specifically, the *g* and its distribution provide a direct measurement of atomic misfit in the HEA. The nm-sized inclusions give rise to precipitation-like strengthening, which has long been a common strengthening method in metals (*40*). The formation of inclusions can be attributed to interaction between the distortion field and chemical inhomogeneity, which plays a significant role in the HEA as our study shows.

**Conclusion**. We have developed a comprehensive approach toward the characterization of lattice distortions in chemically complex alloys by examining local crystal symmetry and mapping of local strain field and paracrystalline distortion parameter. Using this approach, we demonstrate two sources of lattice distortions in the HEA of $Al_{0.1}CrFeCoNi$. First is nm-sized paracrystals and the second features 10±3 nm, disc-shaped, clusters having ~7.1% tensile displacements along <110> directions. The severe distortion in the paracrystals limit their size and contributes to strain between paracrystal blocks. Furthermore, the interaction of nm-sized paracrystal blocks and strained nano-clusters gives rise to fractal strain field that differentiates the structure of the HEA from conventional alloys.

**Acknowledgments:** The authors express many thanks to Profs. Peter Liaw, Robert S. Averback, and Pinshane Huang, Drs. James Mabon, Changqiang Chen, and Xun Zhan and Mr. Haw-Wen Hsiao for the helpful discussions, and Prof. Peter Liaw for providing the HEA sample.

**Funding:** The work is supported by DOE BES (Grant No. DEFG02-01ER45923). RLY is supported by SRC. Electron microscopy experiments were carried out at the Center for Microanalysis of Materials at the Frederick Seitz Materials Research Laboratory of University of Illinois at Urbana-Champaign.

**Author contributions:** Y.T.S. carried out experiments and simulations. Y.T.S. and J.M.Z. analyzed the data and wrote the manuscript with support from all authors. R.Y. contributed to the strain analysis. Y.H. assisted with finite element analysis. Q.Y. contributed to the sample preparation. J.M.Z. supervised the project.

**Competing interests:** Authors declare no competing interests.

**Supplementary Materials:**

Materials and Methods

Figures S1-S6

References

**Supplementary Materials**

**Materials and Methods**

Sample and TEM specimen preparation

Polycrystalline $Al_{0.1}CrFeCoNi$ HEA, provided by Prof. Peter Liaw of University of Tennessee, was prepared by vacuum induction melting and casting (Sophisticated Alloys Inc., Butler, PA). The as-cast samples were hot isostatic pressed at 1,100°C for 1h under a 207 MPa ultra-high-purity argon pressure to reduce porosity, which resulted large grain sizes (~ $10^2$ μm). The composition and chemical homogeneity were checked by atom probe tomography analysis previously (*1*). TEM specimens were prepared along [001] axis from a selected grain in the polycrystalline sample using focused ion beam (FIB) following the procedures described in ref. (*2*). The orientation of the grain was determined with the help of electron backscatter diffraction (EBSD).

Scanning electron nanodiffraction (SEND) acquisition

The SEND experiments were carried out using a Themis Z STEM (Thermo Fisher Scientific) operated at 300kV, and a JEOL 2200FS FEG TEM with an in-column Omega energy filter and operated at 200kV for energy-filtered CBED. Two types of diffraction patterns were recorded. First, for a spot-like diffraction pattern, we used an electron probe of a semi-convergence angle of 1.1 mrad and a probe size of 1.7 nm in FWHM (full-width at half-maximum). The SEND patterns described in the text were acquired using a 100x100 pixel scan over a sample area of 200x200 $nm^2$. The total acquisition time is ~1hr and during this period the sample drift is less than 3 nm. Second, for CBED patterns, we used an electron probe having a semi-convergence angle of 5.5 mrad and probe size of 0.8 nm in FWHM. The SCBED was performed with a 200x200 pixel scan over a sample area of 200x200 $nm^2$. The total acquisition time of ~4 hrs and the overall sample drift is ~10 nm.

X-ray energy dispersive spectroscopy (EDS)

The X-ray EDS area analysis was performed using an electron probe of 0.8 nm in FWHM, with a 180x180 scan over 180x180 $nm^2$ about the same region where SCBED was performed, having specimen tilted ~6° away from [001] zone axis. The specimen thickness is ~186 nm as measured by CBED analysis. Each spectrum was acquired using a four-quadrant FEI Super-X detector for 1s, spanning the total acquisition time of ~11 hrs. For the analysis, the K-α peak signal for each element in each X-ray spectrum was first smoothed using the Savitzky-Golay filter, which was then fitted with a Gaussian peak to obtain the peak height. The peak height was then normalized with respect to total K-α peak heights in each spectrum (Fig. S5). Here, the peak height ratio maps do not represent atomic percent, but only maps out relative concentration fluctuations since there being strong absorption of Al-K X-rays by Ni (*3*).

**Supplementary Text**

Scanning electron nanodiffraction, strain analysis

Strain analysis via SEND is based on Bragg's law by measuring the change in distance between diffracted beams. For each diffraction pattern, the positions of 9 selected beams, which include the center beam and 8 diffracted beams of low-order reflections, were determined. The beam in the diffraction pattern appears as a disk because of the beam's semi-convergence angle (1.1 mrad for SEND). Dynamical diffraction effects on intensity distribution within the disk.

Because of this, diffraction intensity cannot be used to measure the beam position reliably. We used the circular Hough transform method for detecting disk position at subpixel precision (*4*). The method works by applying first the Sobel filter to the SEND pattern to filter out the disk edge. Then, circular Hough transform is applied the filtered SEND pattern, which transform the filtered pattern of edge circles into a pattern of Hough transformation peaks, with each peak marking the position of a detected circle. The peak position was measured by fitting using a Lorentzian peak model. This measurement was applied to all 9 selected beams, and a 2D reciprocal lattice with $\vec{G_1}$ and $\vec{G_2}$ as the basis vectors was then determined from the measured disk positions. Strain for a given lattice plane of $\vec{G} (= i\vec{G_1} + j\vec{G_2})$ is calculated by

$$\frac{1/|\vec{G}| - 1/|\vec{G_0}|}{1/|\vec{G_0}|}, \quad (1)$$

where $1/|\vec{G_0}|$ is the reference d-spacing for the lattice plane of $\vec{G}$, which was obtained using the calibrated camera constant. The calibration was performed using a standard Si specimen.

Scanning electron nanodiffraction, distortion analysis

The integral breadth was measured by non-linear least-squares fitting (using the Python Scipy package) of the intensity profile taken across the (200), (400), and (600) reflections in the recorded nanodiffraction patterns. The intensity profile was taken across the center of (h00) reflections and averaged over a width of 3 pixels (an example is shown in Fig. 4A in Main Text). The fitting was done using a peak model consisted of a pseudo-Voigt function and an exponential background. Instrumental broadening effect was calibrated using a Si single crystal (Fig. S4). The distortion parameter was obtained by a linear fit of integral breadths for three consecutive diffracted peaks.

Simulations for convergent beam electron diffraction patterns

For a perfect crystal, the CBED simulation is based on the Bloch-wave method, using the neutral atomic scattering factors of Doyle & Turner (*5*), absorption parameters of Bird & King (*6*), a total of 29 beams and 3209 incident beam directions within the CBED disc. For a distorted crystal, the CBED simulation is based on the Bloch-wave scattering matrix method, using the same parameters as above. The crystal potential with lattice displacements ($\vec{R}$) was constructed using the deformable ion approximation

$$U(\vec{r}) = \sum_g U_g \exp\left(-2\pi i \vec{g} \cdot \vec{R}(\vec{r})\right) \exp(2\pi i \vec{g} \cdot \vec{r}) \quad (2)$$

where $U_g$ is the Fourier coefficient of the perfect crystal potential. To simulate CBED patterns from the imperfect crystal of the HEA, we considered a crystal column, which is separated into blocks. Within each block, $U_g$ and $\vec{R}$ are constant. The electron wave function in reciprocal space at thickness t, $\Psi = (\Phi_o(t), \Phi_g(t), \cdots)^T$ (T for transpose), is related to the incident wave $\Psi_o = (1, 0, \cdots)^T$ through

$$\Psi = S_n S_{n-1} \cdots S_1 \Psi_o \quad (3)$$

where $S_n$ stands for the scattering matrix of the nth block of the imperfect crystal and $S$ can be calculated using the Bloch wave method for perfect crystals using

$$S = C \Upsilon C^{-1} \quad (4)$$

Here C is the Bloch wave eigenvector matrix and $\Upsilon$ is the diagonal matrix

$$\Upsilon = \left\{ \exp\left(2\pi i \gamma^i \Delta z\right) \right\} \quad (5)$$

with Δz as the block thickness.

To examine the effects of lattice distortions on the symmetry of simulated CBED patterns, we employed two models for the displacement field $\vec{R}(\vec{r})$. The displacement caused by the elliptical inclusion was approximated with a displaced block of 10 nm in a column of crystal (Fig. 2B), and the magnitude and the direction of $\vec{R}$, and the depth position of the displaced block ($z_R$) were the simulation parameters that were adjusted for a fit to the experimental pattern. In regions free of inclusions, the crystal column was divided into blocks of same thickness. Each block is displaced by ($R_x$, $R_y$, $R_z$), which were randomly generated and followed a Gaussian distribution of width $\sigma_R$ and zero mean (Fig. 2F). The simulations were repeated multiple times for a range of block size ($L_b$) and the standard deviation $\sigma_R$ of displacements, with $L_b$= 1-10 nm and $\sigma_R$ = 0.013-0.023 Å, respectively. The simulated CBED patterns were quantified for their 4-fold rotation symmetry and compared with the experimental data. The results are summarized in Fig. S1.

**Supplementary Figures**

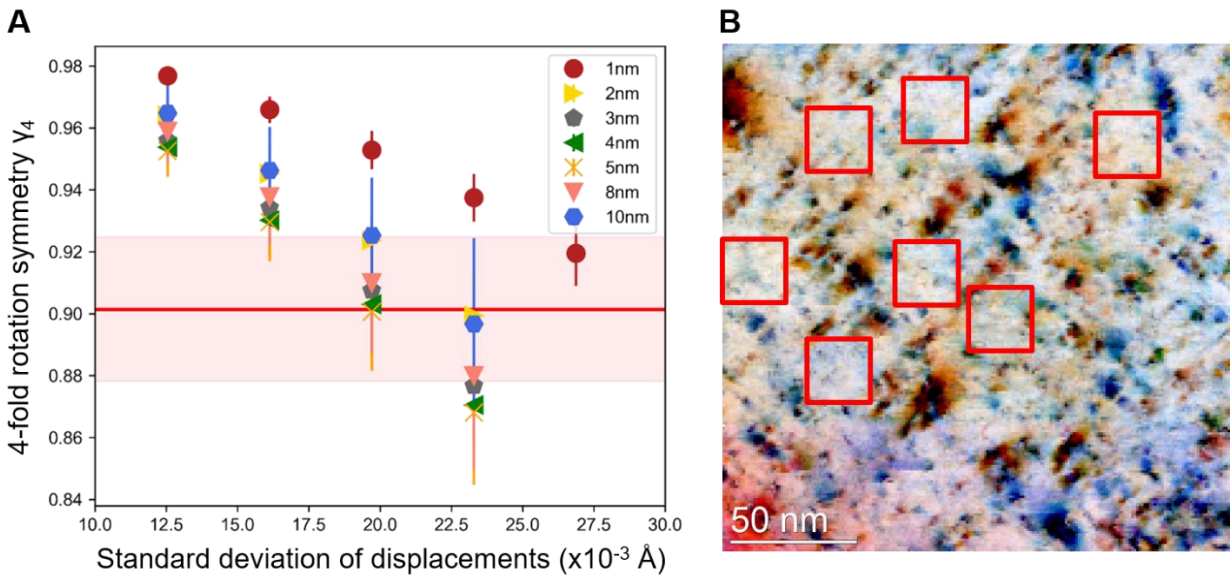

**Fig. S1. Simulations of CBED symmetry fluctuations in Al$_{0.1}$CrFeCoNi using the mosaic crystal models of nm-sized blocks.** (A) The symbols represent the distribution of the measured 4-fold rotational symmetry in the simulated CBED patterns. Each symbol and its bar correspond to the mean and the distribution of 4-fold rotational symmetry from 625 simulated CBED patterns for a given mosaic block size and the specific standard deviation of Gaussian distributed random displacements. The red horizontal line and the band corresponds to the mean and standard deviation of 4-fold rotational symmetry of 4375 experimental CBED patterns from regions free of inclusions, as marked by red boxes in (B). The standard deviation of strain as measured by SEND for (220) and (2$\bar{2}$0) directions are 0.58% and 0.49%, which gives a standard deviation of the directional displacement of 0.006Å and 0.007 Å, respectively. Using these values, the mosaic block size is estimated as between 3-8 nm from Fig. A.

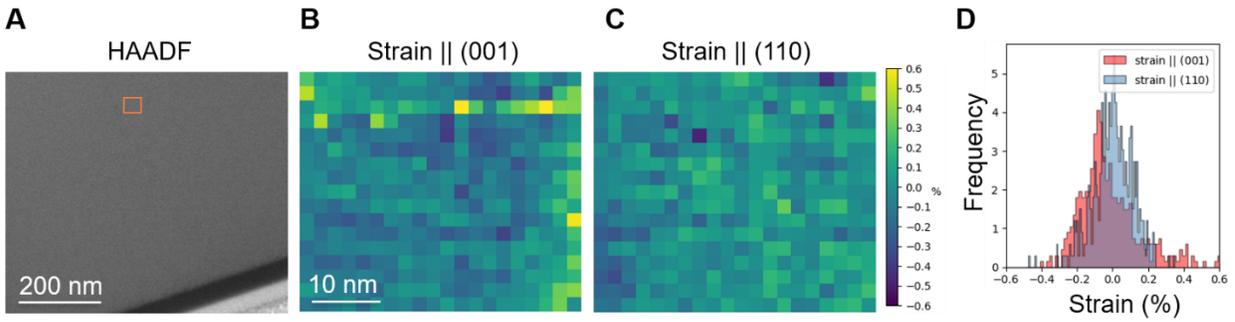

**Fig. S2. Strain** measurement **calibration using a standard Si sample along [110] orientation.**
(A) HAADF image acquired with illumination semi-convergence angle α of 0.7 mrad, inner
collection angle of 60mrad. Orange box indicates the region of the SEND experiment was
performed. Strain maps along (B) (001) and (C) (1$\bar{1}$0) directions were obtained from a 20x17 scan
over 40x34 nm$^2$ using an electron probe of 1.7nm in FWHM. (D) Histogram of strain along two
directions. The thickness of Si sample is selected as similar to the thickness of the measured HEA
sample.

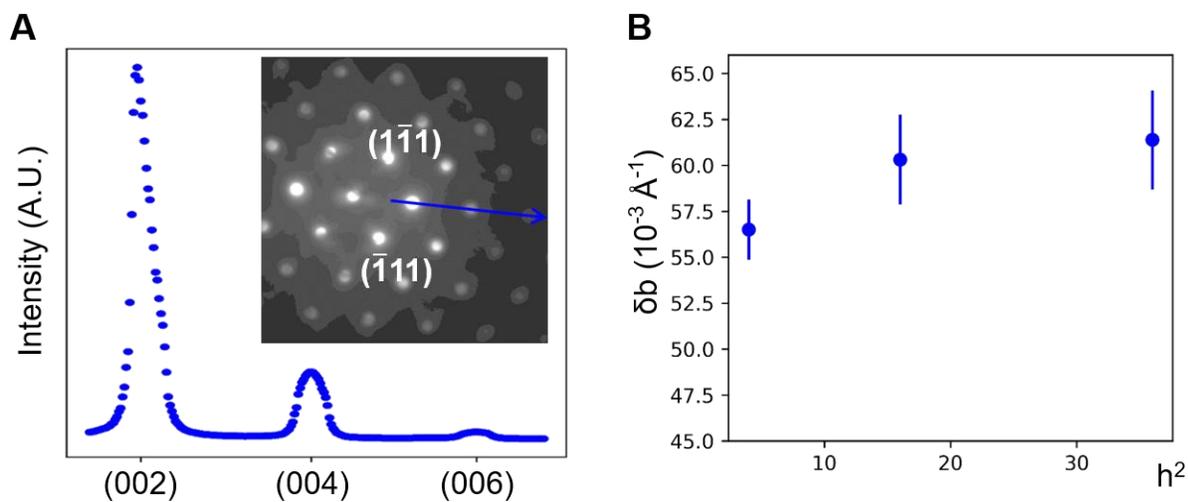

**Fig. S3. Diffraction peak broadening calibration using scanning electron nanodiffraction.** SEND calibration experiment was performed on a standard Si sample along [110] zone axis with the similar experimental conditions as for HEAs. (A) Example of an intensity profile with three orders of (200) reflections drawn from the averaged electron nanodiffraction pattern. (B) Diagram of integral width $\delta b$ versus square of reflection order $h^2$. The error bars represent the standard deviation of 400 diffraction patterns. The narrower peak width for (200) reflections in Si is affected by the intensity oscillations due to strong dynamical diffraction effects.

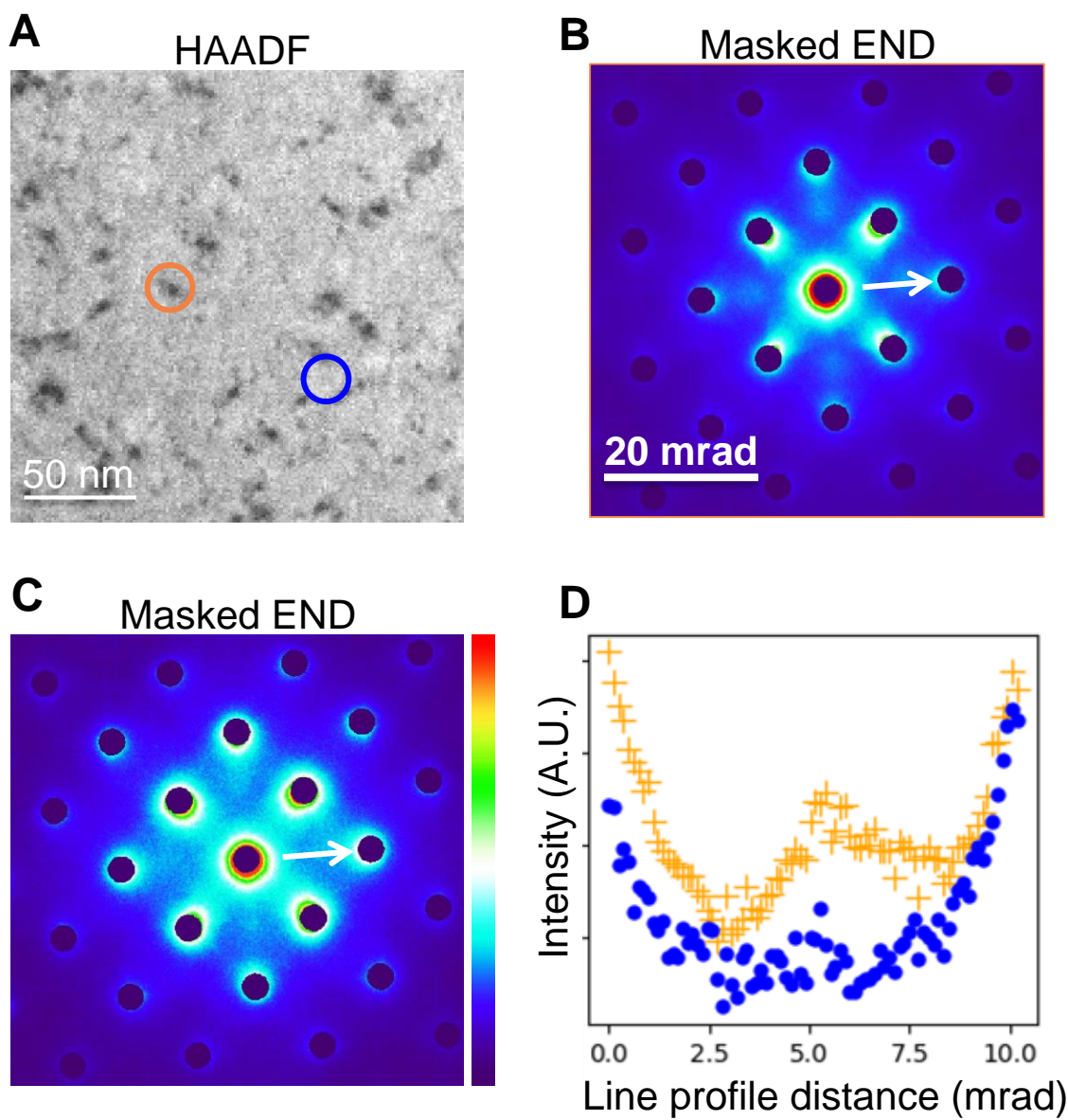

**Fig. S4. Electron nanodiffraction (END) patterns showing strong diffuse scattering.** (A) HAADF image acquired with illumination semi-convergence angle α of 5.5mrad, inner collection angle of 60mrad, corresponding to region of SEND experiment for strain measurements. (B and C) Electron nanodiffraction patterns corresponding to regions marked as orange and blue circles in (A), respectively. The Bragg reflections belonging to an fcc structure were masked off, showing strong diffuse scattering but no extra sharp peaks. (D) Line profiles drawn from the white arrows in (B and C).

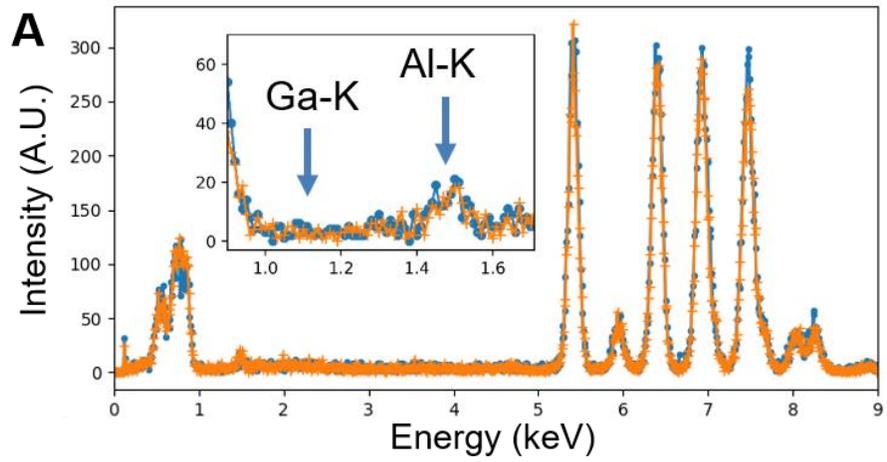

**A**

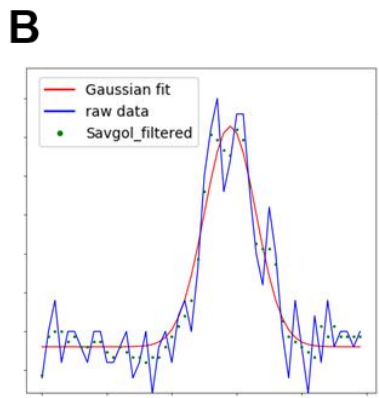

**B**

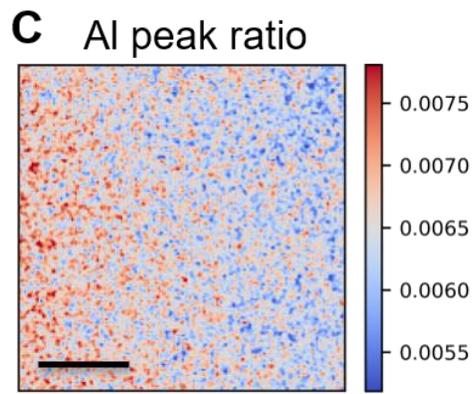

**C** Al peak ratio

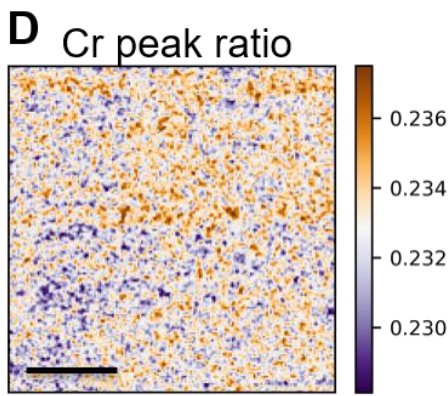

**D** Cr peak ratio

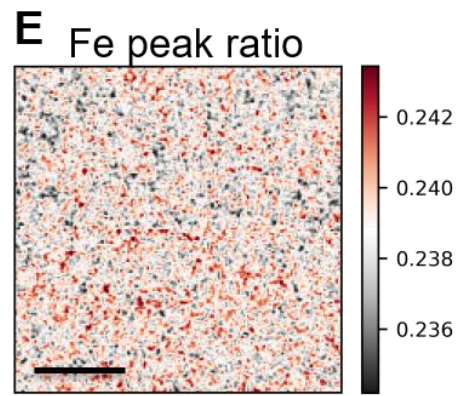

**E** Fe peak ratio

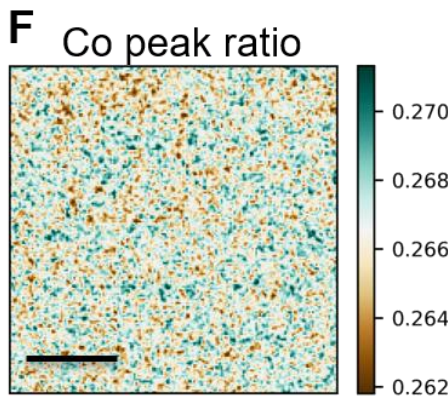

**F** Co peak ratio

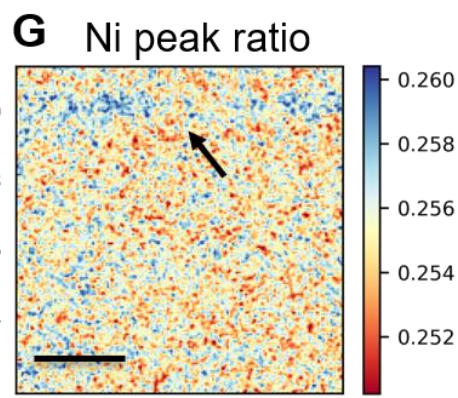

**G** Ni peak ratio

**Fig. S5. Energy dispersive X-ray spectrum images in Al$_{0.1}$CrFeCoNi HEA.** The EDS-spectrum image, 180x180 with step sizes of 1nm, was acquired about the same region as for SCBED, with specimen tilted ~6° away from [001] zone axis. The sample thickness is ~186nm determined by CBED analysis. Each spectrum was acquired for 1s, spanning the total acquisition time of ~11hrs. (A) Two example EDS spectra showing fluctuations in K-α peak heights. The inset shows a zoom in the energy range for the expected Al-K and Ga-K peak positions (the Ga is expected from the preparation of the TEM sample using FIB. The absence of Ga peaks demonstrates that the amount of Ga is less than the detection limit). (B) Selected Al-K peak indicated in (A). Every K-α peak signal for each element was first smoothed with Savitzky–Golay filter, fitted with a Gaussian peak to obtain the peak height, then normalized with respect to total K-α peak heights in each spectrum. Each image was displayed with limits between ±3 times of standard deviation about the average. The scale bar corresponds to 50 nm. The arrow in (G) indicates the horizontal region being damaged by the electron beam due to long exposure.

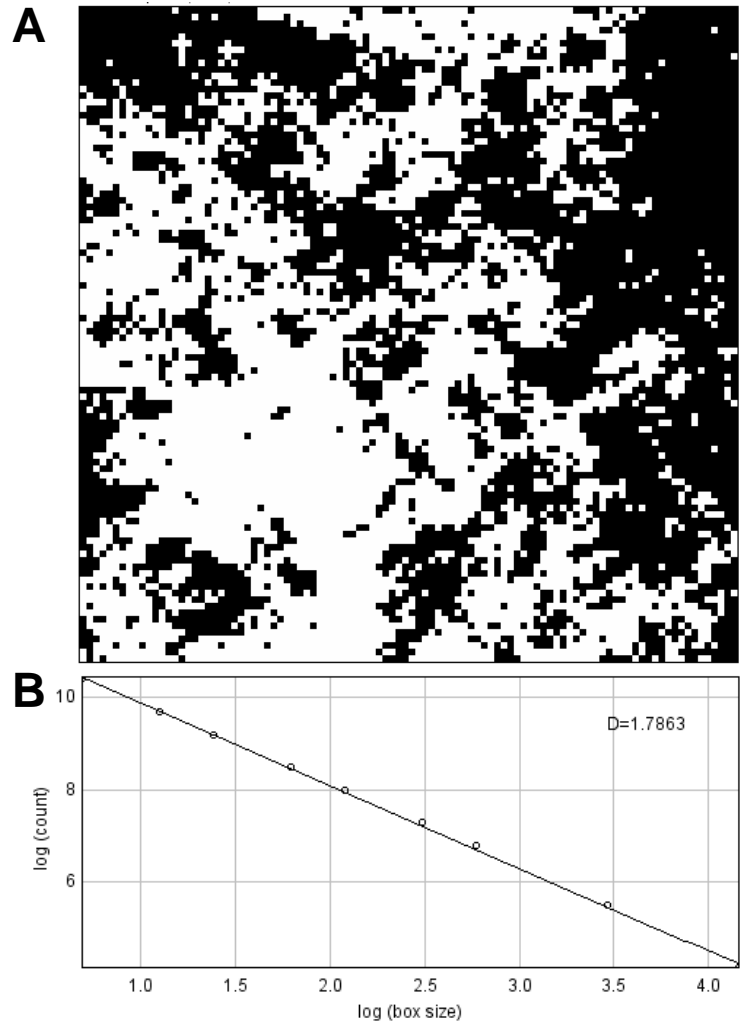

**Fig. S6. Fractal dimension of strain map**. (A) The binary strain map after thresholding (1 for $\varepsilon$ > 0 and 0 else). The strain map has a dimension of 200 nm. (B) The measurement of fractal dimension D using the box counting method as implemented in ImageJ (https://imagej.nih.gov/ij/).